\UseRawInputEncoding\documentclass[aps,pra,floatfix,twocolumn,superscriptaddress,showpacs,10pt]{revtex4-1}
\usepackage{amssymb, amsmath,color,mciteplus,graphicx,subfigure}
\usepackage{calc,epsfig,epstopdf,color,mciteplus,bm,mathrsfs}
\usepackage{times}
\usepackage{float}
\usepackage{lipsum}
\usepackage{verbatim}
\begin{document}

\title{Nonadiabatic Holonomic Quantum Computation via Path Optimization}
\author{Li-Na Ji}
\author{Yan Liang}
\author{Pu Shen}
\affiliation{Guangdong Provincial Key Laboratory of Quantum Engineering and Quantum Materials, and School of Physics\\ and Telecommunication Engineering,  South China Normal University, Guangzhou 510006, China}

\author{Zheng-Yuan Xue} \email{zyxue83@163.com}
\affiliation{Guangdong Provincial Key Laboratory of Quantum Engineering and Quantum Materials, and School of Physics\\ and Telecommunication Engineering,  South China Normal University, Guangzhou 510006, China}

\affiliation{Guangdong-Hong Kong Joint Laboratory of Quantum Matter, and Frontier Research Institute for Physics,\\ South China Normal University, Guangzhou  510006, China}

\date{\today}

\begin{abstract}
Nonadiabatic holonomic quantum computation (NHQC) is implemented by fast evolution processes in a geometric way to withstand local noises. However, recent works of implementing NHQC are sensitive to the systematic noise and error. Here, we present a path-optimized NHQC (PONHQC) scheme based on the non-Abelian geometric phase, and find that a geometric gate can be constructed by different evolution paths, which have different responses to systematic noises. Due to the flexibility of the PONHQC scheme,  we can choose an optimized  path that can lead to excellent gate performance. Numerical simulation shows that our optimized scheme can greatly outperform the conventional NHQC scheme, in terms of both fidelity and robustness of the gates. In addition, we propose to implement our strategy  on superconducting quantum circuits with decoherence-free subspace encoding with the experiment-friendly two-body exchange interaction. Therefore, we present a flexible NHQC scheme that is promising for the future robust quantum computation.
\end{abstract}

\maketitle

\section{INTRODUCTION}
Quantum computation \cite{Nielsen} represents an alternative paradigm that utilizes fundamental principles of quantum mechanics to speed up computational tasks. The promise of quantum computation lies in the possibility of efficiently handling some hard tasks, such as quantum simulation \cite{feymann},  factorization of large prime numbers \cite{Shor},  large-scale searching \cite{grover}, etc. In this paradigm, a set of  universal quantum gates are the building block. Accurate manipulation requires quantum gates featuring two issues, i.e., high fidelity and strong robustness of the implemented quantum gates. 

Since geometric phases do not rely on the details of evolution path but their global properties, a geometric quantum gate possesses intrinsic tolerance to certain local noises. Thus, to fight against noises and/or errors, quantum computation based on geometric phases is a natural and promising strategy. According to the types of geometric phase, namely Abelian \cite{Abelian} and non-Abelian \cite{nonAbelian}, quantum computation can be classified as geometric \cite{GQC, GQCZSL2002, UnGQCZSL, NGQCPZZ, NGQCCT, NGQCCT2020, DWZ} and holonomic quantum computation (HQC) \cite{HQC, AHQC, AHQC_DLM, NHQC1, NHQC2, environment, environment2}, respectively. As the non-Abelian geometric phase is in the matrix form, HQC is naturally considered to be capable of achieving a set of universal quantum gates. Early HQC schemes were realized by the adiabatic evolution \cite{AHQC, AHQC_DLM},  which have been demonstrated in experiments \cite{ExpAHQC, ExpAHQC2}. However, the adiabatic process needs a long run time, which implies slow quantum gate operation, resulting in more infidelity of the implemented quantum gate, due to the inevitable decoherence effect. Subsequently, for eluding the adiabatic limitation, nonadiabatic holonomic quantum computation (NHQC) was proposed \cite{NHQC1, NHQC2} and generalized \cite{SL, SL0, SL1, SL2, SL3,  PZZ, LY, NHQC3, ZJ18, ChenAn, bjliu, ShenPu, LN, CYH, ZJ} to speed up the process while holding the robustness advantage. Later, it was demonstrated in various experimental systems such as superconducting circuits \cite{super1, super2, super3, super4}, nuclear magnetic resonance \cite{NMR, SLexper2, ZhuZN}, nitrogen-vacancy centers in diamond \cite{NV1, NV2, NV3, SLexper1, SLexper3, NV4}, etc.

In geometric and holonomic quantum computation, different evolution paths may lead to the same geometric phase. In  the traditional single-loop NHQC scheme \cite{SL, SL0, SL1, SL2}, the evolution path is enclosed by two longitude geodesics, and it  can realize arbitrary single-qubit holonomic quantum gates. Meanwhile, the shortest-path NHQC scheme \cite{LY} based on a round evolution path also can accomplish the same gates as that of the single-loop NHQC scheme. However, the noise resistance varies greatly with different evolution paths \cite{ABpath1, ABpath2, DingCY}, and thus, for the same gate, different evolution paths have different performance in terms of gate fidelity and robustness. Accordingly, it is worthwhile to analyze the different paths that can obtain the same desired gate, focusing on enhancing the overall gate performance. However, when implementing a geometric quantum gate, previous works set a fixed path in the first place, and thus have no such degree of freedom  to balance  gate fidelity and robustness.

In this work, a NHQC scheme with path optimization to enhance the overall  gate performance  is proposed, where the noises can be greatly suppressed by picking a suitable path.  Firstly, we propose  the scheme and illustrate its implementation in $\Lambda$-type three-level quantum systems. Secondly, proceeding from the unitive path-type that along the longitude and latitude line on Bloch sphere, through numerical simulations, we find that different paths have different pulse areas, gate fidelity, and noise robustness. Therefore, by choosing a well-behaved path, the implemented gates can have excellent performance, i.e., higher gate fidelity and stronger gate robustness. Finally, we verify the feasibility of our scheme in a superconducting quantum circuit system. In addition, to suppress the collective dephasing noise, we adopt the decoherence-free subspace (DFS) \cite{DFS1, DFS2, DFS3} encoding, achieving  arbitrary single-qubit gates and nontrivial two-qubit gates, with conventional two-body exchange interactions between qubits. Therefore, our scheme provides a clue to select an evolution path featuring outstanding gate performance, and is a promising endeavor to achieve high-fidelity and strong-robustness quantum gates in large-scale quantum computation.

\section{GENERAL METHOD OF GEOMETRIC PATH DESIGN}
\subsection{Holonomic quantum computation}
First, we consider a quantum system governed by a general Hamiltonian $\mathcal{H}(t)$, there exists an $N$-dimensional Hilbert subspace spanned by a complete set of the basis vectors $\{|\Psi_k(t)\rangle_{k=0}^{N-1}\}$, where the time-dependent states $|\Psi_k(t)\rangle$ evolve according to the Schr\"{o}dinger equation $\textrm{i}|\dot{\Psi}_k(t)\rangle\!=\!\mathcal{H}(t)|\Psi_k(t)\rangle$. Thus, the corresponding time-evolution operator driven by the Hamiltonian $\mathcal{H}(t)$ can be obtained as
\begin{equation}
\label{U}
U(t)=Te^{-\textrm{i}\int_0^t \mathcal{H}(t') \textrm{d}t'}=\sum_{k}|\Psi_k(t)\rangle \langle \Psi_k(0)|,
\end{equation}
where $T$ is the time-ordering operator. Here, to derive our target geometric phase \cite{NHQC1, A-Aphase}, we introduce a set of auxiliary bases $\{|\psi_k(t)\rangle\}$ that satisfy the boundary condition of cyclic evolution at the initial moment $t\!=0$ and the final moment $t=\tau$, i.e., $|\psi_k(0)\rangle\!=\!|\psi_k(\tau)\rangle\!=\!|\Psi_k(0)\rangle$. In this way, we can expand  $|\Psi_k(t)\rangle=\sum_lc_{kl}(t)|\psi_l(t)\rangle$. Substitute  it into the Schr\"{o}dinger equation, then we get
\begin{equation}
\dot c_{kl}(t)={\rm i}\sum_m c_{km}(t)[{\rm i}\langle \psi_m(t)|\dot\psi_l(t)\rangle-\langle \psi_m(t)|\mathcal{H}(t)|\psi_l(t)\rangle].
\end{equation}
At the final time, the evolution operator can be written as
\begin{eqnarray}
\label{Utau}
U(\tau)=T\sum_{m,l}e^{-{\rm i}[A_{ml}(\tau)+K_{ml}(\tau)]}|\psi_m(0)\rangle\langle\psi_l(0)|,
\end{eqnarray}
where
\begin{eqnarray}
\label{phaseAK}
A_{ml}(\tau)&\equiv&{\rm i}\int_0^\tau\langle \psi_m(t)|\dot\psi_l(t)\rangle {\rm d}t, \notag\\
K_{ml}(\tau)&\equiv&-\int_0^\tau\langle \psi_m(t)|\mathcal{H}(t)|\psi_l(t)\rangle {\rm d}t
\end{eqnarray}
are the elements of  holonomic matrix and dynamical matrix, respectively. Generally, the dynamical phase always impinges on the noise-resilient feature, so the total phase we set depends only on geometric features by ensuring $K_{ml}=\eta A_{ml}$, where $\eta$ is a proportional constant independent on the parameters of qubit system. On the one hand, by eliminating the dynamical element, i.e., $\eta\!=\!0$, the total phase will be a pure geometric phase. On the other hand, if $\eta\!\neq\!0, -1$, it reduces to an unconventional geometric phase \cite{UnGQCZSL}.

\begin{figure}[tbp]
  \centering
  \includegraphics[width=\columnwidth]{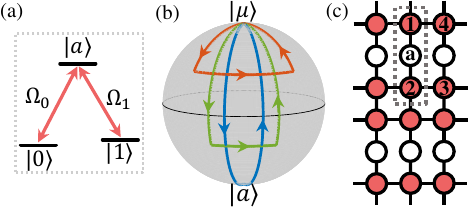}
 \caption{Illustrations of the proposed scheme. (a) The $\Lambda$-type three-level structure, microwave fields are applied to induce the resonant transitions from two-qubit states $|0\rangle$, and $|1\rangle$ to the auxiliary level $|a\rangle$. (b) Illustration of different paths for the same holonomic quantum gate on the Bloch sphere. The blue path is the conventional single-loop scheme. (c) Schematic diagram of scalable 2D superconducting qubit lattice. The solid (hollow) circles represent transmon (cavity) physical qubits, respectively. The encircled part by the rectangle is the DFS encoded logical qubit, which includes two transmons connected by an auxiliary cavity.}\label{shiyitu}
\end{figure}

\subsection{Application to the $\Lambda$-type three-level structure}
We now consider the $\Lambda$-type three-level structure, as shown in Fig. \ref{shiyitu}(a). Microwave fields drive the transition from a target qubit level $|j\rangle$ $(j=0,1)$ to auxiliary level $|a\rangle$ with coupling strength $\Omega_j(t)$, phase $\phi_j(t)$ and the same detuning $\Delta(t)$, and the corresponding Hamiltonian will be
\begin{eqnarray}
\label{Ho}
\mathcal{H}(t)\!=\!\Delta(t)|a\rangle\langle a|\!+\!\frac {1} {2}\left[\sum_{j=0,1}\Omega_j(t)e^{-\textrm{i}\phi_j(t)}|j\rangle\langle a|\!+\!{\rm H.c.}\right],
\end{eqnarray}
we set $\Omega_0(t)\!=\!\Omega(t)\sin(\theta/2)$, $\Omega_1(t)\!=\!\Omega(t)\cos(\theta/2)$, and $\phi\!=\!\phi_0(t)-\phi_1(t)$, where $\theta$ and $\phi$ are adjustable parameters. A set of orthogonal auxiliary bases are selected as follows:
\begin{eqnarray}
\label{Estates}
|\psi_0(t)\rangle\!&=&\!\cos\frac{\theta}{2}e^{-{\rm i}\phi}|0\rangle\!-\!\sin\frac{\theta}{2}|1\rangle, \notag\\
|\psi_1(t)\rangle\!&=&\!\cos\frac{\chi(t)}{2}|\mu\rangle\!+\!\sin\frac{\chi(t)}{2}e^{{\rm i}\xi(t)}|a\rangle, \notag\\
|\psi_2(t)\rangle\!&=&\!\sin\frac{\chi(t)}{2}e^{-{\rm i}\xi(t)}|\mu\rangle\!-\!\cos\frac{\chi(t)}{2}|a\rangle,
\end{eqnarray}
where $|\mu\rangle\!=\!\sin(\theta/2)|0\rangle+\cos(\theta/2)e^{{\rm i}\phi}|1\rangle$. Particularly, $|\psi_0(t)\rangle$ decouples from the dynamics of $\mathcal{H}(t)$. In addition, for satisfying the boundary condition of cyclic evolution: $|\psi_k(0)\rangle\!=\!|\psi_k(\tau)\rangle\!=\!|\Psi_k(0)\rangle$, we need to ensure $\chi(0)=\chi(\tau)=0$, and this implies that $|\psi_2(0)\rangle$ is not in the computational subspace. Therefore, we take the $|\psi_1(t)\rangle$ as an example, on which the global phase $\gamma$ is accumulated at the final time. As shown in Fig. \ref{shiyitu}(b), the detailed evolution path of $|\psi_1(t)\rangle$ is shown on the Bloch sphere by visualized parameters $\chi(t)$ and $\xi(t)$, which represent the polar and azimuth angles, respectively, in the range of $[0,\pi]$ and $[0,2\pi)$. Meanwhile, by solving the Schr\"{o}dinger equation, the parameter-limited relationships can be confirmed as
\begin{equation}
\begin{split}
\label{limitedP}
\dot{\chi}(t)&=\Omega(t)\sin[\phi_0(t)-\xi(t)],\\
\dot{\xi}(t)&=-\Delta(t)-\Omega(t)\cot\chi(t)\cos[\phi_0(t)-\xi(t)].
\end{split}
\end{equation}
Therefore, once our target evolution path dominated by $\chi(t)$ and $\xi(t)$ is settled, we can further design Hamiltonian parameters $\Omega(t)$ and $\phi_0(t)$ reversely. In addition, an unitary evolution operator can be obtained following Eq. (\ref{U}) as
\begin{eqnarray}
\label{unGeoU}
U(\tau)&=&|\Psi_0(0)\rangle \langle \Psi_0(0)|+e^{{\rm i}\gamma}|\Psi_1(0)\rangle \langle \Psi_1(0)| \notag\\
       &=&\exp\left({\rm i}\frac{\gamma}{2}\right)\exp\left(-{\rm i}\frac{\gamma}{2} \vec{{\rm n}}\cdot\vec{\sigma}\right)
\end{eqnarray}
within the computational subspace $\{|0\rangle, |1\rangle\}$, which is equivalent to a rotation operation of angle $\gamma$ around the $\vec{\textrm{n}}\cdot\vec{\sigma}$ axis, where $\vec{\textrm{n}}\!=\!(-\sin\theta \cos\phi, -\sin\theta \sin\phi, \cos\theta)$, and $\vec{\sigma}=(\sigma_x, \sigma_y, \sigma_z)$ are the Pauli operators. Thus, the $X-$, $Y-$ and $Z-$axes rotation gates for arbitrary angle $\gamma$, denoted as $R_{x,y,z}(\gamma)$, can be implemented by setting $(\theta, \phi)=(\frac{\pi}{2}, \pi)$, $(\frac{\pi}{2}, -\frac{\pi}{2})$, and $(0, \forall)$, respectively. It is worth emphasizing that the choice of the different rotation axes just relate to $(\theta, \phi)$ not to $(\chi, \xi)$, so there is no additional restriction of path parameters $(\chi, \xi)$ in constructing different gates.

Different evolution paths have different responses to systematic error and noises. However, the previous schemes result in a fixed geometric path choice due to the rigorous geometric conditions, so that the path is inevitably located on the one that is more sensitive to systematic error and noise, resulting in the loss of qubit information. Therefore, we give the solutions to unfreeze the limitations so that the evolution paths can be selectable. In order to obtain a geometric evolution, we analyze the global phase $\gamma$ accumulated on the $|\psi_1(\tau)\rangle$ for example. The global phase is the sum of two parts, with
\begin{eqnarray}
\label{phasegd}
\gamma_g\!=\!A_{11}&\!=\!&-\frac{1}{2}\int^{\tau}_0\dot{\xi}(t)[1\!-\!\cos\chi(t)]{\rm d}t,\notag\\
\gamma_d\!=\!K_{11}&\!=\!&\frac{1}{2}\int^{\tau}_0\frac{\dot{\xi}(t)\sin^2\chi(t)\! +\!\Delta(t)[1\!-\!\cos\chi(t)]}{\cos\chi(t)}{\rm d}t,
\end{eqnarray}
where $\gamma_g$ as the geometric phase is exactly half of the solid angle enclosed by the path, and $\gamma_d$ is the dynamical phase. Furthermore, for achieving the path-optimized NHQC (PONHQC), we can treat the dynamical phase in the  following two ways.

Firstly, we strictly eliminate the dynamical phase $\gamma_d\!=\!0$ and obtain a pure geometric phase $\gamma\!=\!\gamma_g$. In this case, the detuning parameter needs to meet $\int_0^\tau\Delta(t){\rm d}t=-\int_0^\tau\dot{\xi}(t)[1+\cos\chi(t)]{\rm d}t$, only in this way the path parameters $(\chi, \xi)$ can be flexible for optimization. Otherwise, from Eqs. (\ref{limitedP}) and (\ref{phasegd}), additional constraint $\phi_0(t)=\xi(t)\pm\pi/2$ needs to be met to ensure $\gamma_d=0$ in the case of $\Delta(t)=0$, leading to a fixed orange-slice-shaped path as the blue line shown
in Fig. \ref{shiyitu}(b), thus no alternative path for the purpose of optimization.

Secondly, we turn to implement an unconventional geometric phase \cite{UnGQCZSL} by a simple resonant interaction $\Delta=0$. Set $\gamma_d=\eta\gamma_g$, where $\eta$ is a constant that is dependent on the $\chi$, then the total phase
\begin{eqnarray}
\label{gamma}
\gamma\!=\!(1\!+\!\eta)\gamma_g =\!-(1\!+\!\eta)\frac{1}{2}\!\int^{\tau}_0\!\!\dot{\xi}(t)[1\!-\!\cos\chi(t)]{\rm d}t
\end{eqnarray}
is an unconventional geometric phase. For a holonomic gate operation in Eq. (\ref{unGeoU}) under a target rotation angle $\gamma$, we can find in Eq. (\ref{gamma}) that path parameter $\chi(t)$ [or $\xi(t)$] still has different choices while satisfying the cyclic evolution boundary conditions [$\chi(0)\!=\!\chi(\tau)\!=\!0$]. That is,   a same gate can be realized by driving on different evolution paths.

To sum up, the path-optimized purpose can be realized by the above two ways, i.e., with or without detuning. However, removing the dynamical phase requires a subtle choice of control parameters or more operations than that needed in the dynamical process, which increases the complexity  experimentally. Therefore, we adopt the unconventional geometric way.

We consider a set of evolution paths along the longitude and latitude lines on the Bloch sphere as an example, which is the extension of the conventional orange-slice-shaped path, as shown in Fig. \ref{shiyitu}(b), thus the comparison results between the optimized and conventional paths can be clear at a first glance. In terms of coordinates $(\chi, \xi)$, the path undergoes
\begin{equation}
A(0,\xi_1)\rightarrow B(\chi,\xi_1)\rightarrow C(\chi, \xi_2)\rightarrow A(0,\xi_2),
\end{equation}
that is, the path first starts from the north pole $A(0,\xi_1\!)$ and evolves along the longitude line with $\xi(t)\!=\!\xi_1$ to point $B(\chi,\xi_1)$ at the time $\tau_1$; then evolves along the latitude line with $\chi(t)\!=\!\chi$ to point $C(\chi, \xi_2\!)$ at time $\tau_2$; and returns to the north pole $A(0,\xi_2\!)$ at the final time $\tau$ along the longitude line with $\xi(t)\!=\!\xi_2$. Therefore, according to the parameter-limited relationships in Eq. (\ref{limitedP}), we can solve the Hamiltonian parameters $\Omega(t)$ and $\phi_0(t)$ corresponding to these three path segments $t\in[0, \tau_1)$, $[\tau_1, \tau_2)$ and $[\tau_2, \tau]$ as
\begin{eqnarray}
\label{3path+}
\int^{\tau_1}_0\Omega(t){\rm d}t\!&=&\!\chi, \quad\quad\quad\ \  \phi_0(t)\!=\xi_1+\frac{\pi}{2}, \notag\\
\int^{\tau_2}_{\tau_1}\Omega(t){\rm d}t\!&=&\!2\gamma\cot\frac{\chi}{2}, \ \ \phi_0(t)\!=\pi\!+\!\cot\chi\int^{t}_{\tau_1}\Omega(t'){\rm d}t', \notag\\
\int^{\tau}_{\tau_2}\Omega(t){\rm d}t\!&=&\!\chi, \quad\quad\quad\ \ \phi_0(t)\!=\xi_2-\frac{\pi}{2},
\end{eqnarray}
respectively. The above $\gamma$ is considered the case of $\gamma\!>0$, and when $\gamma\!<0$, the second segment in Eq. (\ref{3path+}) should be changed to $\!\int^{\tau_2}_{\tau_1}\Omega(t){\rm d}t\!=\!-2\gamma\cot(\chi/2)$ and $\phi_0(t)\!=\!-\cot\chi\int^{t}_{\tau_1}\Omega(t'){\rm d}t'$. Note that, for this simple configuration of obtaining the geometric phase, there is no restriction on the shape of $\Omega(t)$, i.e., it can be arbitrary. As shown in Fig. \ref{shiyitu}(b), there are different geometric paths by setting $\chi$ to vary in the range of $[0,\!\pi]$, or setting $\xi_2-\xi_1$ to vary in the range of $[0,2\pi)$, for a target rotation angle $\gamma\!=\!-(1\!+\!\eta)(\xi_2-\xi_1)(1\!-\!\cos\chi)/2$, in which $\eta\!=\!-(1\!+\!\sec\chi)$. And, to ensure that the unconventional geometric condition is met, we should keep the path parameter $\chi$ (and $\eta$) consistent during the construction of a set of universal holonomic gates once it is considered as the suitable path.

\begin{figure*}[tbp]
  \centering
\includegraphics[width=\textwidth]{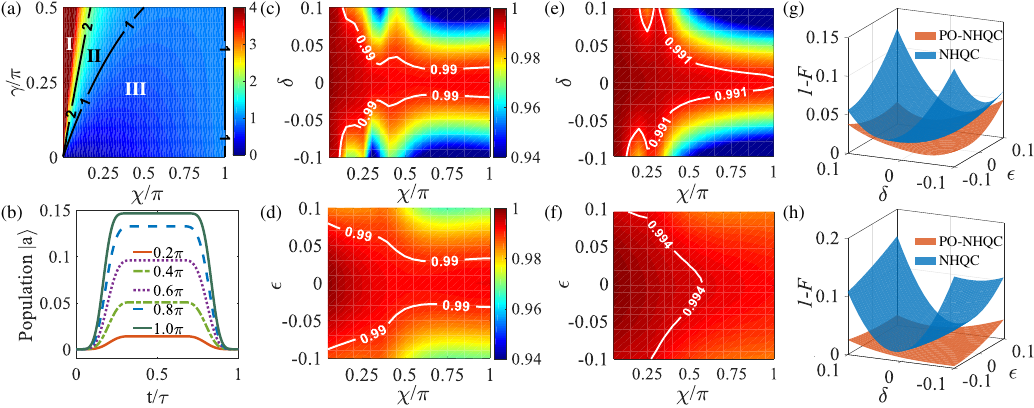}
  \caption{Numerical simulation results for single-qubit gates. (a)  The pulse area $\mathcal{S}\!=\!\int_0^\tau \Omega(t){\rm d}t/2$ in units of $\pi$, as functions of the  path parameter $\chi$ and the rotation angle $\gamma$, where small pulse area means short evolution time. In part I, $\mathcal{S}\!>\!2\pi$; in part II, $\pi\!<\!\mathcal{S}\!<\!2\pi$; and in part III, $0\!<\!\mathcal{S}\!<\!\pi$. (b) The state population dynamics of the auxiliary state $|a\rangle$ for different paths. As $\chi$ decreases, the shape is squashed.  (c) and (d) The fidelity of $R_x(\pi/2)$ gate as functions of $\chi$ and $\delta$, and $\chi$ and $\epsilon$, respectively. And, (e) and (f) are the same work as (c) and (d) for the $R_y(\pi/4)$ gate. (g) and (h) are the robustness test results for $R_x(\pi/2)$ and $R_y(\pi/4)$ gates, respectively. We simulate the gate infidelities $1\!-\!F$ as functions of  $\delta$ and $\epsilon$ simultaneously, the surfaces of our PONHQC scheme with $\chi=\pi/4$ (orange surface) have gentler bends than that of the conventional NHQC  scheme with $\chi=\pi$ (blue surface).}\label{simulations}
\end{figure*}

\subsection{The gate performance}
In this subsection, based on the above path type in $\Lambda$-type three-energy structure, we show how to pick out a satisfying evolution path, then the gate performances are tested based on this path. Systematic errors destroy the conditions of geometric cyclic evolution, and greatly weigh the gate performance down. Therefore, singling out a path that withstands systematic errors can further strengthen geometric gates robustness. Here, we consider the detuning error $\delta$ and Rabi error $\epsilon$ induced by imperfect control based on the ideal Hamiltonian, in the form of $\delta\Omega_\textrm{m}|a\rangle\langle a|$ and $[(\epsilon \Omega_0/2)|0\rangle\langle a|e^{-{\rm i}\phi_0}+(\epsilon \Omega_1/2)|1\rangle\langle a|e^{-{\rm i}\phi_1}]+{\rm H.c.}$, with a simple pulse shape $\Omega(t)\!=\!\Omega_\textrm{m}\sin^2(\pi t/\tau)$. Meanwhile, evolution time has considerable difference for different $\chi$ and cannot be overlooked. Therefore, we explore how the pulse area $\mathcal{S}\!=\!\int_0^\tau \Omega(t){\rm d}t/2$ changes with $\chi$ and $\gamma$ in Fig. \ref{simulations}(a): for orange-slice-shaped path ($\chi=\pi$), it is always $\pi$ pulse no matter how big the rotation angles are, for $\chi<\pi$, larger angle rotations take a longer time for the same path $\chi$; in part III, $\mathcal{S}<\pi$, and part I reads $\mathcal{S}>2\pi$, so we have to fully analyze the decoherence factor caused by the long time exposure to the environment. The system suffering the decoherence impact is modeled by the master equation of
\begin{equation}
\label{density}
\dot\rho(t)=-{\rm i}[\mathcal{H}(t),\rho(t)]+\sum_{i=-,z}\frac{\kappa_i}{2}\mathcal{L}(D_i),
\end{equation}
where $\rho$ is density operator, and $i\!=\!-, z$ is to distinguish decay and dephasing operator, respectively. $\kappa_i$ represent the decay and dephasing rates.  $\mathcal{L}(D_i)\!=\!2D_i\rho D_i^{\dagger}\!-D_i^{\dagger}D_i\rho\!-\rho D_i^{\dagger}D_i$ is the Lindblad operator. In simulation, we consider the decay and dephasing operators of
$D_{-}=|0\rangle\langle a|+|1\rangle\langle a|$, $D_{z}=2|a\rangle\langle a|-|1\rangle\langle 1|-|0\rangle\langle 0|$, 
and the decoherence scale of $\kappa_{-}\!=\!\kappa_{z}\!=\!\Omega_{\rm m}/2000$. In addition, we find the population of auxiliary state $|a\rangle$ is depressed when $\chi$ decreases, as shown in Fig. \ref{simulations}(b). This phenomenon drops a hint that the smaller $\chi$ possesses the intrinsic resistant to the transitions of excited state, i.e., resistant to the effects of $D_-$ and $D_z$. This will be one of the cores for selecting an optimized path.

Another consideration for the optimized path is the performance of gates. We use definition of the gate fidelity \cite{GF} of $F\!=\!(1/2\pi)\int_0^{2\pi}\langle\psi_f|\rho(\tau)|\psi_f\rangle{\rm d}\theta_1$ to evaluate the quality of single-qubit gates, in which $|\psi_f\rangle$ is the ideal final state obtained by $|\psi_f\rangle\!=\!U(\tau)|\psi_i\rangle$, and $\rho(\tau)$ is the imperfect density operator comes out of Eq. (\ref{density}). The gate fidelity is the average result for a general initial states $|\psi_i\rangle\!=\!\cos\theta_1|0\rangle+\sin\theta_1|1\rangle$, and the $\theta_1\in[0, 2\pi]$ is the traversal factor signifying 1000 different initial inputs. We take two noncommutative gate operations $R_{x}(\pi/2)$ and $R_{y}(\pi/4)$ gates as typical examples. In Figs. \ref{simulations}(c) and (d) for $R_{x}(\pi/2)$ gate and Figs. \ref{simulations}(e) and (f) for $R_{y}(\pi/4)$ gate, we simulate the gate fidelity $F$ influenced by $\chi$ with the  errors $\delta, \epsilon$ disturbing at $[-0.1, 0.1]$, which show that different paths have different sensitivity to $\delta$ and $\epsilon$ errors. Generally, it is more preferable to choose smaller $\chi$ for stronger gate robustness. However, smaller $\chi$ also leads to the longer gate time, and thus lower gate fidelity.   Therefore, balancing the fidelity with the robustness, we set the path parameter to be $\chi=\pi/4$ in the following.

we compare the gate robustness between our PONHQC scheme with $\chi=\pi/4$ and  NHQC scheme. We plot the gate infidelity $1\!-\!F$ with different $\delta$, $\epsilon$ errors in Figs. \ref{simulations}(g) and (h) for $R_x(\pi/2)$ and $R_y(\pi/4)$ gates, respectively. One can see the whole orange -surfaces have gentler bends, which means that PONHQC obviously exceeds the conventional NHQC scheme. Indeed, not only the above two gates, other rotation gates $R_{x,y,z}(\gamma)$ with $\gamma\in[0, \pi/2]$ all have high-fidelity and strong-robustness superiorities. Thus, path optimization strategy plays a role in improving the error robustness and gate fidelity of holonomic quantum gates.

\section{Physical implementation}
\subsection{Single-logical-qubit holonomic gates}
In this section, our above theoretical scheme is implemented in the superconducting circuit system to demonstrate the feasibility and necessity of the scheme, with the DFS encoding \cite{ DFS1, DFS2, DFS3}, which can greatly suppress the collective dephasing due to the symmetrical structure of the interaction between the qubits and surrounding environment. There are a pair of transmons connected by an auxiliary cavity, as the encircled part by the rectangle in Fig. \ref{shiyitu}(c), and the Hamiltonian of this system can be written as
\begin{equation}
\begin{split}
\mathcal{H}_{1}=&\sum_{n=1}^{+\infty}\{\sum_{j=1,2}\left[n\omega_j-\frac{n(n-1)}{2}\alpha_j\right] |n\rangle_j\langle n|+n\omega_a|n\rangle_a\langle n|\\
&+[g_{ja}\sqrt n|n-1\rangle_j\langle n|\otimes|n\rangle_a\langle n-1|+{\rm H.c.}]\},\\
\end{split}
\end{equation}
where $\omega_{j(a)}$ is qubit frequency, $\alpha_j$ is the anharmonicity of transmon, and $g_{ja}$ is the coupling strength between transmon and auxiliary cavity. A logical qubit is encoded by two transmons ${\rm T}_1$ and ${\rm T}_2$, i.e.,
\begin{equation}
S_1={\rm Span}\{|0\rangle_L=|10\rangle_{12}, |1\rangle_L=|01\rangle_{12}\}.
\end{equation}
Whereas, qubit parameters, such as frequency and coupling strength, are fixed, not only is it difficult to realize the wanted energy interactions, but it also leads to the degradation of gate performance due to not working in the optimal parameters' area. Hence the experimental demonstrated parametrically tunable coupling technique \cite{Tunable7, Tunable8, Tunable9, Tunable10}, achieved by biasing the transmon with an ac magnetic flux, is absolutely imperative. For this purpose, we add frequency driving on each transmon in the form of $\omega_j(t)=\dot F_j$ $(j=1,2)$ with $F_j(t)=\beta_j\sin[\nu_jt+\varphi_j(t)]$. We set $g_{1a}=g_{2a}=g$. In the interaction picture, the Hamiltonian can be written as
\begin{equation}\label{driven H}
\begin{split}
\mathcal{H}'_{1}=\sum_{j=1,2}g|10\rangle_{ja}\langle 01|e^{-{\rm i}\Delta_j(t)}e^{{\rm i}F_j(t)}+{\rm H.c.},
\end{split}
\end{equation}
in which $\Delta_j=\omega_a-\omega_j$ is the frequency difference of cavity and transmon. The term $e^{{\rm i}F_j(t)}$ can be expanded using Jacobi-Anger identity of
\begin{equation}\label{Jacobi}
\exp({\rm i}z\sin\zeta)=\sum_{n=-\infty}^{+\infty}J_n(z)\exp({\rm i}n\zeta),
\end{equation}
where $J_n(z)$ represent $n$-order Bessel function at $z$. In Eq. ({\ref{driven H}}), resonant transitions of $|10\rangle_{ja}\leftrightarrow|01\rangle_{ja}$ can be achieved by setting $\nu_j=\Delta_j$. After undergoing the rotation wave approximation, the $\mathcal{H}'_{1}$ is truncated to
\begin{equation}
\begin{split}
\mathcal{H}_1^{{\rm eff}}&=\frac{1}{2}\sum_{j=1,2}\Omega_j|10\rangle_{ja}\langle 01|e^{-{\rm i}\varphi_j(t)}+{\rm H.c.}\\
&=\frac{1}{2}[\Omega_1|0\rangle_{L}\langle a|e^{-{\rm i}\varphi_1(t)}+\Omega_2|1\rangle_{L}\langle a|e^{-{\rm i}\varphi_2(t)}]+{\rm H.c.}
\end{split}
\end{equation}
where $\Omega_j=2gJ_1(\beta_j)$ and the auxiliary state being $|a\rangle=|010\rangle_{1a2}$. Consequently, we successfully construct an effective Hamiltonian $\mathcal{H}_1^{{\rm eff}}$ like Eq. (\ref{Ho}) in superconducting system.

\begin{figure}[tbp]
  \centering
  \includegraphics[width=8cm]{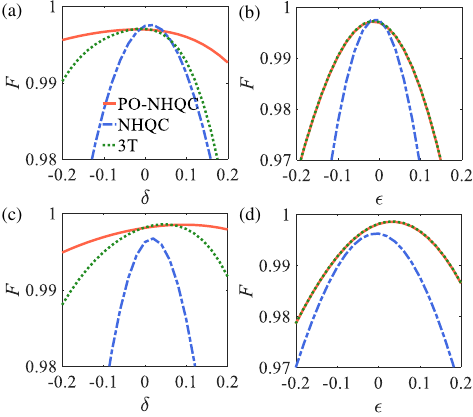}
\caption{Performance of the single-qubit gates using the Hamiltonian in Eq. (\ref{driven H}). (a) and (b) $R_x(\pi/2)$  gate robustness, for our PONHQC scheme and the traditional NHQC scheme, as the functions of   $\delta$ and  $\epsilon$, respectively. In addition, if the auxiliary qubit is replaced by transmon, the corresponding robustness results based on our scheme are labeled by `3T'. (c) and (d) The corresponding results for $R_y(\pi/4)$ gate.}
  \label{Robustness}
\end{figure}

Next, we take into account the imperfect controls occurring in this systems, and further test the robustness. Systematic errors, especially the qubit-frequency drift, is a real headache in superconducting system. In advance, qubit-frequency drift error $\delta$ and the driving amplitude deviation $\epsilon$ are all considered, and the $\delta, \epsilon$-error Hamiltonians are expressed as

\begin{equation}
\begin{split}\label{Herror}
&\mathcal{H}_1^{\delta}=g\delta_1|1\rangle_1\langle1|+g\delta_2|1\rangle_2\langle1|,\\
&\mathcal{H}_1^{\epsilon}=\sum_{j=1,2}g\epsilon_j|10\rangle_{ja}\langle 01|e^{-{\rm i}\Delta_j(t)}e^{{\rm i}F_j(t)}+{\rm H.c.},
\end{split}
\end{equation}
respectively. In simulation, we give the appropriate values as $g\!=\!2\pi\times10$ MHz, $\beta_1\!=\!1.7$ and $\Delta_1\!=\!\Delta_2\!=\!2\pi\times400$ MHz. In addition, we consider the decoherence effect. For transmon-cavity-transmon configuration in Fig. \ref{shiyitu}, we give them footnotes $1, c, 2$ orderly, and the decay/dephasing operators are
\begin{equation}
\label{D1}
D_{1-}\!=\sum_{i=1,c,2}|0\rangle_i\langle1|,\quad D_{1z}\!=\sum_{i=1,2}(|1\rangle_i\langle1|-|0\rangle_i\langle0|).
\end{equation}
From the state-of-the-art experiment \cite{IBM}, we take the decoherence rates of $\kappa_-\!=\!\kappa_z\!=\!2\pi\times3$ KHz. In Fig. \ref{Robustness}, we compare the gate robustness constructed by our PONHQC scheme ($\chi=\pi/4$ as optimized path) and conventional NHQC scheme for $R_x(\pi/2)$ and $R_y(\pi/4)$ gates under the disturbance of $\delta\!=\!\delta_{1,2}$, $\epsilon\!=\!\epsilon_{1,2}\in[-0.2, 0.2]$. Apparently, the PONHQC scheme (red line) possesses stronger robustness than the NHQC scheme (blue line).

In addition, we talk about the effect of physical qubit type for encoding. In the above configuration, we adopt a cavity as the auxiliary qubit to connect two transmons. The advantage is that some error and noise source of cavity can be negligible. We also consider replacing the auxiliary qubit by a transmon to implement our scheme, i.e., transmon-transmon-transmon (3T) way. In the 3T scheme, we consider the decoherence of
\begin{equation}
D_{1-}'\!=\sum_{i=1,t,2}|0\rangle_i\langle1|, \quad
D_{1z}'\!=\sum_{i=1,t,2}(|1\rangle_i\langle1|-|0\rangle_i\langle0|),
\end{equation}
the subscript `$t$' represents the auxiliary transmon qubit. The decay and dephasing rates are $\kappa_-\!=\!\kappa_z\!=\!2\pi\!\times\!3$ KHz. We take the error into account, the $\epsilon$-affected Hamiltonian $\mathcal{H}_1^{\epsilon'}=\mathcal{H}_1^{\epsilon}$ is unchanged, but the $\delta$-affected Hamiltonian turns into
\begin{equation}
\mathcal{H}_1^{\delta'}=g\delta'_1|1\rangle_1\langle1|+g\delta'_t|1\rangle_t\langle1| +g\delta'_2|1\rangle_2\langle1|,
\end{equation}
$\delta\!=\!\delta'_{1,2}\!=\!-\delta'_t\in[-0.2, 0.2]$. The robustness results using 3T configuration are shown in Fig. {\ref{Robustness}}, where the curves are labeled by `3T'. By comparing the red line and green line, we can see the sensitivity to $\epsilon$ error is near for two configurations, but the former configuration has the better robustness advantage than 3T for $\delta$ error, which is the main error source in the superconducting circuit system.

Next, we evaluate gate fidelity the system can arrive. We set $g=2\pi\times 10$ MHz, and optimize the parameters of $\beta_1=1.7$ and $\Delta_1=\Delta_2=2\pi\times390$ MHz by searching the optimal parameter region. In the end, the gate fidelities of $R_x(\pi/2)$ and $R_y(\pi/4)$ can reach 99.78$\%$ and 99.84$\%$, respectively. The gate  infidelity is mainly due to the rotating wave approximation in getting the effective Hamiltonian and the decoherence effect. For $R_x(\pi/2)$ gate, the infidelities from these two sources are 0.03$\%$ and 0.19$\%$, respectively. And, for $R_y(\pi/4)$ gate, the infidelities are 0.06$\%$ and 0.10$\%$, respectively.

\begin{figure}[tbp]
  \centering
  \includegraphics[width=8cm]{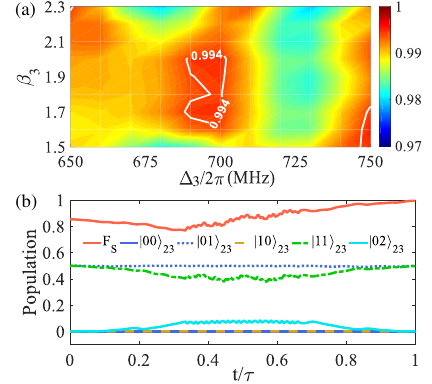}
  \caption{Simulations of two-qubit control-phase gate. (a) Numerical search for the optimal work area of Bessel parameter $\beta_3$ and frequency difference $\Delta_3$, evaluated by gate fidelity. (b) The state populations with the initial state $(|01\rangle_{23}+|11\rangle_{23})/\sqrt2$, at the final time, the state fidelity $F_S$ reads $99.50\%$. }\label{twoqubit}
\end{figure}

\subsection{Two-logical-qubit holonomic gates}

In addition to the above single-qubit gates, a nontrivial two-qubit element is also needed for a set of universal quantum gates, so we next set out to the nontrivial control-phase gate. Using two pairs of transmon qubits, i.e., T$_1$ and T$_2$, T$_3$ and T$_4$ to encode the first and second DFS logical qubits, respectively, the two-logical qubit bases span a four-dimensional DFS, i.e.,
\begin{equation}
\begin{split}
S_2={\rm Span}\{&|1010\rangle_{1234}=|00\rangle_L, |1001\rangle_{1234}=|01\rangle_L, \\
&|0110\rangle_{1234}=|10\rangle_L, |0101\rangle_{1234}=|11\rangle_L\}.\\
\end{split}
\end{equation}
The two logical units are coupled by ${\rm T}_2$ and ${\rm T}_3$, and the coupling strength is $g_{23}$. We add the frequency driving on ${\rm T}_3$ with $\omega_3(t)=\dot F_3$, $F_3(t)=\beta_3\sin[\nu_3t+\varphi_3(t)]$. Similarly, we take the first-order Bessel function, and thus the interaction Hamiltonian of ${\rm T}_2$ and ${\rm T}_3$ can be written as
\begin{equation}
\begin{split}
\mathcal{H}_{23}=&g_{23}J_1(\beta_3)e^{i[\nu_3 t+\varphi_3(t)]}\{|10\rangle_{23}\langle01|e^{-{\rm i}\Delta_3 t}\\
&+\sqrt2|11\rangle_{23}\langle02|e^{-{\rm i}(\Delta_3-\alpha_3)t}\\
&+\sqrt2|20\rangle_{23}\langle11|e^{-{\rm i}(\Delta_3+\alpha_2)t}\}+{\rm H.c.},
\end{split}
\end{equation}
where $\Delta_3\!=\!\omega_3-\omega_2$ is the frequency difference of ${\rm T}_3$ and ${\rm T}_2$. We set $\nu_3=\Delta_3-\alpha_3$ to product the resonant transition of $|11\rangle_{23}\leftrightarrow|02\rangle_{23}$, and the Hamiltonian can be written as
\begin{eqnarray}
\mathcal{H}'_{23}&=&g_{23}J_1(\beta_3)e^{{\rm i}\varphi_3(t)}\{|10\rangle_{23}\langle01|e^{-{\rm i}\alpha_3 t}\\
&&+\sqrt2|11\rangle_{23}\langle02|
+\sqrt2|20\rangle_{23}\langle11|e^{-{\rm i}(\alpha_2+\alpha_3)t}\}+{\rm H.c.}.\notag
\end{eqnarray}
Similar to the  single-qubit gate case, we divide the entire evolution time $\tau$ into three parts at time moment $\tau_1$ and $\tau_2$. In order to accumulate a geometric phase $\gamma$ on the state $|11\rangle_{23}$ at the final time, the pulse areas of three parts need to satisfy
\begin{eqnarray}
\label{2bit}
&\Omega'\tau_1=\chi;\ \Omega'(\tau_2-\tau_1)=\!2\gamma\cot\frac{\chi}{2};\ \Omega'(\tau-\tau_2)\!=\!\chi,
\end{eqnarray}
where $\Omega'\!=\!2\sqrt2g_{23}J_1(\beta_3)$, and the phase $\varphi_3$ can be set arbitrarily. Then within the computational space $S_2$, the control-phase gate operation ${\rm CP}(\gamma)\!=\!$ diag$(1, 1, 1, e^{{\rm i}\gamma})$ can be formed.

Next, we evaluate the two-logical-qubit CP($\pi/4$) gate, and the two-qubit gate fidelity is defined by $F_2=(1/4\pi^2)\int_0^{2\pi}\int_0^{2\pi}\langle\psi_f|\rho(\tau)|\psi_f\rangle {\rm d}\theta_1{\rm d}\theta_2$, where the ideal final state is $|\psi_f\rangle\!=\!{\rm CP}(\gamma)|\psi_i\rangle$, and the initial state being the direct product state $|\psi_i\rangle\!=\!(\cos\theta_1|10\rangle_{12}+\sin\theta_1|01\rangle_{12}) \otimes(\cos\theta_2|10\rangle_{34}+\sin\theta_2|01\rangle_{34})$, in which $\theta_1, \theta_2\in[0, 2\pi]$ are traversal
factors signifying different 10 000 initial inputs. The density operator $\rho(\tau)$ is the output of master equation, in which we consider the decay and dephasing operators of
\begin{equation}
\begin{split}
D_{2-}&=\sum_{i=1}^4(|0\rangle_i\langle1|+\sqrt2|1\rangle_i\langle2|),\\
D_{2z}&=\sum_{i=1}^4(|1\rangle_i\langle1|+2|2\rangle_i\langle2|),\\
\end{split}
\end{equation}
and the corresponding rates are $\kappa_-\!=\!\kappa_z\!=\!2\pi\times3$ KHz. Set parameters $g_{23}=2\pi\times8$ MHz, $\alpha_2=2\pi\times300$ MHz, $\alpha_3=2\pi\times330$ MHz, and we optimize the parameter $\beta_3$ and qubit frequency difference $\Delta_3$ numerically, as shown in Fig. \ref{twoqubit}(a). Accordingly, we pick $\beta_3=2$ and $\Delta_3=2\pi\times700$ MHz as the appropriate optimization parameters. Under the above settings, the gate fidelity of the ${\rm CP}(\pi/4)$ can be as high as $99.50\%$. In addition, state populations with the initial state $(|01\rangle_{23}+|11\rangle_{23})/\sqrt2$ are shown in Fig. \ref{twoqubit}(b), from which we can see the transition process between the states $|11\rangle_{23}\leftrightarrow|02\rangle_{23}$, and the leakage of $|11\rangle_{23}$ is little at the final time. The corresponding state fidelity $F_S=\langle\psi_f|\rho(\tau)|\psi_f\rangle$ is $99.50\%$ at the final time. The gate infidelities due to the rotational wave approximation and the decoherence effect are 0.20$\%$ and 0.30$\%$, respectively.

\section{DISCUSSION AND CONCLUSION}
In conclusion, we propose the path-optimized NHQC scheme to solidify the holonomic gate performance, using the  unconventional geometric phase. By exploring a set of paths numerically, we find that different paths hold quite different behaviors, like pulse area, gate fidelity and robustness, according that we can pick out a satisfying path. In addition, as we do not set the used pulse shape, our proposed scheme can be compatible to various optimal control techniques, which can further enhance the performance of the quantum operations. In physical implementation, we prove the feasibility of the above path-optimized scheme in a superconducting quantum circuit system, with DFS encoding to suppress the collective dephasing error. Consequently, a path-optimized NHQC scheme is feasible and can obtain better performance than the traditional NHQC scheme. The gate fidelities of single-qubit gates are about $99.80\%$ and two-qubit control-phase gate is $99.50\%$.

\acknowledgements
This work is supported by the Key-Area Research and Development Program of GuangDong Province (Grant No. 2018B030326001), the National Natural Science Foundation of China (Grant No. 11874156), and Guangdong Provincial Key Laboratory (Grant No. 2020B1212060066).

\end{document}